\documentclass[letter]{aa}
\usepackage{txfonts}
\usepackage{amssymb}
\usepackage{amsmath}
\usepackage{subcaption}
\usepackage{gensymb}
\usepackage{wasysym}
\usepackage{graphicx}
\usepackage{braket}

\usepackage{txfonts}

\begin{document}

\title{Collisional polarization of molecular ions: a signpost of ambipolar diffusion}
\titlerunning{Collisional polarization of HCO$^+$}
\author{Boy Lankhaar
        \and
        Wouter Vlemmings
        }
\institute{Department of Space, Earth and Environment, Chalmers University of Technology, Onsala Space Observatory, 439 92 Onsala, Sweden \\
\email{boy.lankhaar@chalmers.se}}

\date{Received ... ; accepted ...}

\abstract
   {Magnetic fields play a role in the dynamics of many astrophysical processes, but they are hard to detect. In a partially ionized plasma, a magnetic field works directly on the ionized medium but not on the neutral medium, which gives rise to a velocity drift between them: ambipolar diffusion. This process is suggested to be important in the process of star formation, but has never been directly observed. We introduce a method that could be used to detect ambipolar diffusion and the magnetic field that gives rise to it, where we exploit the velocity drift between the charged and neutral medium. By using a representative classical model of the collision dynamics, we show that molecular ions partially align themselves when a velocity drift is present between the molecular ion and its main collision partner H$_2$. We demonstrate that ambipolar diffusion potently aligns molecular ions in regions denser than their critical density, which subsequently leads to partially polarized emission from these species. We include a model for HCO$^+$ and show that collisional polarization could be detectable for the ambipolar drifts predicted by numerical simulations of the inner protostellar disk regions. The polarization vectors are aligned perpendicular to the magnetic field direction projected on the plane of the sky.}


\maketitle
\section{Introduction}
Ambipolar diffusion arises in a partially ionized and magnetized plasma, where the neutral and ionized medium are collisionally coupled \citep{mestel:56, lizano:89, fiedler:93}. The magnetic force that works on the charged medium causes the ions and electrons to drift with respect to the neutral components of the plasma. The velocity drift is determined by the balance of the magnetic force and the friction between the charged and neutral medium. Through this friction, the magnetic field thus couples indirectly to the neutral medium. Ambipolar diffusion is generally thought to be a prime regulatory agent in the process of star formation \citep{mouschovias:99}. However, ambipolar diffusion has yet to be confirmed through direct observations \citep{yen:18}.

The primary characteristic of ambipolar diffusion is a velocity drift between the neutral and charged medium of an astrophysical plasma. The magnitude of this drift velocity determines the dynamical influence of ambipolar diffusion. At large scales, an order-of-magnitude estimate of the drift velocity between the ionized and neutral medium is $0.85\ (n/10^6\ \mathrm{cm}^{-3})^{0.61}$ km~s$^{-1}$ \citep{draine:10}. Closer to the protostar, the velocity slip is expected to be higher, but estimates of the drift velocity on smaller scales vary greatly. Numerical magnetohydrodynamic (MHD) simulations predict the ambipolar drift to be  $\sim 0.1-1$ km~s$^{-1}$ at hundreds of au from the central protostar \citep{ciolek:98, li:11, zhao:18}. ALMA observations by \citet{yen:18} were unable to show such a velocity drift between H$^{13}$CO$^+$ and $^{18}$CO in their line profiles toward B335. Instead, \citet{yen:18} constrained the ambipolar drift to be smaller than $0.3$ km~s$^{-1}$ at scales of $100$ au. 

Because it is on the order of the thermal velocity of the molecular components that make up the gas in a star-forming region, a velocity drift below $0.3$ km~s$^{-1}$ is hard to detect, even with the spectral resolution of ALMA. In this Letter, we show that when ambipolar drift velocities are on the order of thermal velocities, a significant alignment manifests itself in a linear molecular ion, such as HCO$^+$, through preferentially directed collisions with the neutral medium. The molecular alignment subsequently leads to a partial linear polarization of the emitted radiation, which will be a signpost of both ambipolar diffusion and the magnetic field direction. 

It has long been known that atoms tend to collisionally align themselves in beam expansion experiments, and emit polarized radiation as a result \citep{ellett:26}. Investigating collisions between partially aligned atoms through the polarized emission that they emit after the scattering event has very successfully elevated our understanding of atomic collisions \citep{andersen:01}. Molecules also align themselves in molecular beam experiments \citep{friedrich:91b}. Molecular alignment can be attained through the interaction with a modest electric field \citep{friedrich:91}, but even without an electric field, partial alignment in the population of linear molecules arises naturally as the result of a velocity difference between the expansion carrier gas and the molecule of interest \citep{sinha:74, friedrich:91b}.

Although it is analogous to mechanical alignment, one of the possible dust alignment mechanisms \citep{gold:52, lazarian:97}, the alignment of molecules through collisions has, as far as we have been able to find, not been the subject of any astrophysical investigations. Rather, the emergence of polarization in thermal molecular lines is commonly thought to be quenched by collisions \citep{goldreich:81, lankhaar:20}. Alignment will not manifest itself when collisions are randomly oriented, which is the case for the majority of molecular collisions in astrophysical regions. However, through the process of ambipolar diffusion, a velocity drift arises between the ionized and the neutral medium that leads to a preferred direction of collisions. In this Letter, we show that these (partially) directional collisions lead to detectable polarization.

\section{Collisional polarization}
{\bf Molecular alignment:} We considered the collisional interaction between a linear molecular ion and an H$_2$ molecule in the (spherically symmetric) $j=0$ state. It is usual to treat ion-neutral collisions using the Langevin model, where the collisional dynamics are described by the charge-dipole interaction \citep{draine:10}. The Langevin cross section captures mainly elastic collisions, which transfer linear momentum between the collision partners. We did not follow this approach because we are mostly interested in inelastic collisions---governed by an anisotropic potential \citep{flower:99}---that change the magnitude and orientation of the angular momentum vector around which the molecular ion rotates. We took the anisotropy in the potential to be represented by a solid-body collision between the HCO$^+$-H$_2$ complex. We treated the collision and angular momentum character of the molecular ion classically. We represent the molecular ion by an ellipsoid, characterized by the major- and minor-axis parameters, $a'$ and $b'$, and we let H$_2$ be a sphere of radius $r$. If we define $a=a'+r$ and $b=b'+r$, then the cross section of a collision between the collision partners is given by \citep{sinha:74} 
\begin{align}
\sigma_{\Theta} = \pi ba \sqrt{\left(\frac{b}{a}\right)^2 \sin^2 \Theta +  \cos^2 \Theta }, 
\end{align} 
where $\cos \Theta = \hat{\boldsymbol{k}} \cdot \hat{\boldsymbol{v}}_{\mathrm{rel}}$ is the angle between the direction of the relative velocity of the collision complex, $\hat{\boldsymbol{v}}_{\mathrm{rel}}$, and the orientation of the molecular ion, $\boldsymbol{k}$, at the time of the collision. The relative velocity of the collision complex is simply the velocity difference between the neutral H$_2$ and the molecular ion. It is a function of both the directional drift velocity, $\boldsymbol{v}_{\mathrm{drift}}$ and the randomly oriented thermal velocities, $\boldsymbol{v}_{\mathrm{thermal}}$, of the collision partners. 

\begin{figure}[h!]
  \centering
  \includegraphics[width=0.5\textwidth]{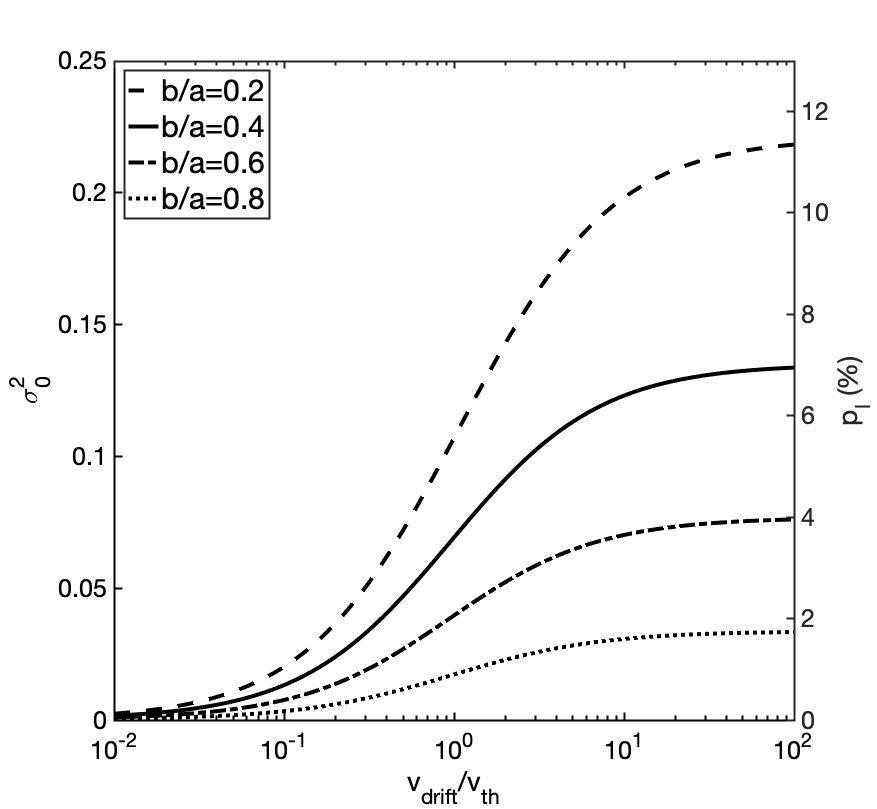}
  \caption{Relative alignment of the molecular states as a function of the ratio, $v_{\mathrm{drift}}/ v_{\mathrm{th}}$, of the ambipolar drift velocity to the thermal velocity for different anisotropy-parameters $b/a$. On the right-hand axis we note the associated predicted optically thin polarization fraction of a $J=3-2$ transition (see Eq.~\ref{eq:thin_pol}). For the HCO$^+$-H$_2$ complex, we assume $b/a=0.4$.}
  \label{fig:rel_align}
\end{figure}

We focused on `hard collisions', which transfer a relatively large amount of angular momentum \citep{sinha:74}. In a hard collision, the collisional product has no memory of its original angular momentum direction and randomizes the product angular momentum orientation. In this picture, the rate at which the molecular ion, oriented with respect to the ambipolar diffusion direction, $\hat{\boldsymbol{z}}$, by angle $\cos \theta = \boldsymbol{k}\cdot \hat{\boldsymbol{z}}$, is scattered to a new angle $\theta'$, is
\begin{align}
k_{\theta \to \theta'} &= \frac{1}{2} n_{\mathrm{H}_2} v_{\mathrm{drift}} \sigma_\theta + \frac{1}{2} \braket{n_{\mathrm{H}_2} \sigma_0 v} =\frac{k_0}{2} \left[\frac{\sigma_{\theta}}{\sigma_0} \frac{v_{\mathrm{drift}}}{v_{\mathrm{thermal}}} + 1 \right], 
\end{align}  
where the factor $2$ is a normalization factor. We let $n_{\mathrm{H}_2}$ be the molecular hydrogen number density, and $k_0 = \braket{n_{\mathrm{H}_2} \sigma_0 v}$ is the thermal collision rate. The thermal collision rate is dependent on the angularly averaged collision rate $\sigma_0 = \frac{1}{2}\int_{-1}^{1} d\cos \theta \ \sigma_\theta$ because of the random orientation of the thermal velocities. When we consider the time-dependence of the population of molecular ions at the orientation $\theta$, $n_{\theta}$, we have
\begin{align}
\dot{n}_{\theta} = -n_{\theta} \int d\cos \theta' \ k_{\theta \to \theta'} + \int d\cos \theta' \ n_{\theta'} k_{\theta' \to \theta}. 
\label{eq:state_align}
\end{align} 
Because the collisional timescale is significantly shorter than the dynamical timescale, we can assume steady state: $\dot{n}_{\theta}=0$. Under the assumption of steady state, Eq.~(\ref{eq:state_align}) can be solved under the physical constraint, $\frac{1}{2}\int_{-1}^{1} d\cos \theta \ n_{\theta} = n$. The relative alignment of the molecular states is $\left[\sigma_0^2\right]^{\mathrm{drift}} = \sqrt{5}\braket{P_2 (\cos \theta) n_{\theta}/n}$ \citep{blum:81}. The square brackets denote that alignment is defined with respect to the velocity drift direction. The magnetic precession rate is higher than the rate of collisions: $\frac{g\Omega}{k_0} \sim 10^{3} \frac{B/\mathrm{mG}}{n_{\mathrm{H}_2}/10^6 \ \mathrm{cm}^{-3}}$, so the alignment will reorient to the magnetic field after each collision event. The relation between the alignment with respect to the magnetic field and the drift velocity is $\left[\sigma_0^2\right]^{\mathrm{B}} = -\frac{1}{2} \left[\sigma_0^2\right]^{\mathrm{drift}}$, where we described the $90^o$ rotation between the drift-frame and the magnetic field-frame with the Wigner D-matrix element $D_{00}^{(2)}(\pi / 2) = -\frac{1}{2}$. We have thus obtained the molecular alignment relevant to the production of polarized emission: $\left[ \sigma_0^2 \right]^{\mathrm{B}}=\sigma_0^2$. The molecular alignment is marginally dependent on the angular momentum state. We plot the relative alignment of the molecular states, $\sigma_0^2$, as a function of $v_{\mathrm{drift}} / v_{\mathrm{thermal}}$ and for different anisotropy coefficients, $b/a$, in Figure~\ref{fig:rel_align}.

{\bf Emergent polarization:} The emergence of polarization from the partially aligned molecular ions can be computed from the polarized radiative transfer equation \citep{landi:84}. We consider a transition $J' \to J$ at frequency $\nu_0$ and local thermal equilibrium. We formulate the polarized radiative transfer equations in terms of the optical depth
\begin{align}
\tau_{\nu} = \frac{h \nu}{4\pi} B_{JJ'} N_J\left(1 - e^{-h\nu_0/kT}\right) \phi_{\nu} ,
\end{align} 
where $B_{JJ'}$ is the Einstein B-coefficient, $\phi_{\nu}$ is the line profile as a function of the frequency $\nu$, and $N_J$ is the column density for lower level $J$. The polarized radiative transfer coefficients under the assumption of local thermal equilibrium are \citep{landi:06, lankhaar:20} 
\begin{subequations}
\begin{align}
t_i = \frac{\tau_I}{\tau_{\nu}} &= 1 + \sigma_0^2\left[w_{JJ'}^{(2)} - w_{J'J}^{(2)}e^{-h\nu_0/kT} \right]\frac{3 \cos^2 \chi - 1}{2\sqrt{2}}, \\ 
t_q = \frac{\tau_Q}{\tau_{\nu}} &= -\sigma_0^2\left[w_{JJ'}^{(2)} - w_{J'J}^{(2)}e^{-h\nu_0/kT} \right]\frac{3\sin^2 \chi }{2\sqrt{2}}, \\
e_i B_{\nu} (T) = \frac{\epsilon_I}{\kappa_{\nu}} &= B_{\nu} (T)\left[1 + \sigma_0^2 w_{J'J}^{(2)}\frac{3 \cos^2 \chi - 1}{2\sqrt{2}}\right], \\ 
e_q B_{\nu} (T) = \frac{\epsilon_Q}{\kappa_{\nu}} &= -B_{\nu} (T) \sigma_0^2 w_{J'J}^{(2)}\frac{3 \sin^2 \chi}{2\sqrt{2}}, 
\end{align}
\end{subequations}
where $B_{\nu} (T)$ is the Planck function, $w_{JJ'}^{(2)}$ are functions dependent on the angular momentum state of the upper and lower level \citep[see, e.g.,][]{landi:06}, $\kappa_{\nu}$ is the opacity, related to the optical depth through $d\tau_{\nu}=\kappa_{\nu} ds$, and $\chi$ is the angle between the radiative propagation and the magnetic field. It is convenient to express the radiative transfer equation in terms of the parallel and perpendicular polarization components of the radiation field: $I_{\parallel,\perp} = \frac{1}{2}[I \pm Q]$. The polarization direction is with respect to the magnetic field direction projected on the plane of the sky. We define the propagation coefficients in these terms as $t_{\parallel,\perp} = t_i \pm t_q$ and $e_{\parallel, \perp} = (e_i \pm e_q)/2$. We thus note the radiative transfer equation for the polarized components of the radiation field 
\begin{align}
\frac{d}{d\tau_{\nu}} I_{\parallel,\perp}= -t_{\parallel,\perp} I_{\parallel,\perp} + e_{\parallel, \perp} B_{\nu} (T).
\label{eq:pol_rad}
\end{align}
The polarization fraction is defined as equation (10) of \citet{goldreich:81} and can be readily evaluated assuming a background radiation field, which we took as the cosmic microwave background radiation field. We plot the predicted polarization fraction of the $J=3-2$ transition of HCO$^+$ as a function of the optical depth for a number ambipolar drift-thermal velocity ratios in Figure~\ref{fig:pol_frac}. We report the polarization fractions assuming a propagation angle of $\chi=90^o$; for other angles, it is a good approximation to multiply the estimates by $\sin^2 \chi$. The polarization coming from ambipolar drift collisions is directed perpendicular to the projected magnetic field direction. 

In Figure~\ref{fig:pol_frac} we plot the polarization fraction of HCO$^+$ for different drift velocities as a function of the optical depth. We assumed an anisotropy parameter of $b/a=0.4$ for HCO$^+$-H$_2$ collisions in this figure (see the paragraph on model assumptions below for the discussion of this parameter). A large fraction of polarization emerges in the radiation when the drift velocity is greater than the thermal velocity ($v_{\mathrm{drift}} / v_{\mathrm{thermal}} > 10$). At low optical depths our method estimates a polarization fraction of $7 \%$, which increases to $9\%$ for high optical depths. At drift velocities on the order of the thermal velocity ($v_{\mathrm{drift}} / v_{\mathrm{thermal}} \sim 1$), polarization fractions on the order of $4-5 \%$ are expected. When drift velocities are $\text{ten}$ times lower than the thermal velocity, the polarization fraction is estimated to be $0.7-1 \%$. Polarization fractions of $\sim 0.5\%$ arise for $v_{\mathrm{drift}} / v_{\mathrm{thermal}}=0.05$. 

{\bf Model assumptions:} A number of simplifications were made in order to derive the above results. First, we represent the cross section of the H$_2$-molecular ion collision-complex as a classical sphere-prolate ellipsoid collision. We stress that this is a rather simplified modeling of the collision dynamics between a molecular ion and a hydrogen molecule. However, because we have assumed dominant collisions, we only need to capture the relative anisotropy of the inelastic collisions. We estimate the anisotropy factor $b/a=0.4$ for H$_2$-HCO$^+$, which we base on the geometry of the collision complex. It is possible that this anisotropy factor is an overestimation because we did not include elastic collisions. Even though elastic collisions do not change the magnitude of the angular momentum vector, a fraction of them might change its orientation. On the other hand, the orientation of the scattered angular momentum vector is not likely to be random because the collisions have a preferred direction. A more thorough analysis requires quantum-dynamical calculations of the state-to-state differential cross sections. Such cross sections are quantum-state transition specific and might vary in directional character between each other. It is expected that the ortho-to-para ratio of H$_2$ affects the state-to-state differential cross-sections and also that the gas temperature has influence on the cross-section anisotropy. Having state-to-state differential cross sections also allows for the accurate treatment of the quantum angular momentum dynamics of the system, which we assume to behave classically in our current approach. Even though we do not capture the complex collision dynamics, we nevertheless expect that our classical model captures the order of the anisotropy of the collision complex, and thus it provides realistic predictions of the emergent polarization fraction through collisionally induced polarization. 

Second, we assume that the ratio of directional to random collisions is determined by the ratio of the drift velocity to the thermal velocity of the collisional complex. This approach has been successfully employed in molecular beam experiments \citep{friedrich:91b}, but it is only an approximation to the proper incorporation of thermal motions and a drift velocity \citep{alexander:77}. Additionally, it has been shown that the presence of a magnetic field perpendicular to a flow leads to changes in the velocity distribution of both the molecular ions and neutrals \citep{pinto:12}. These are second-order effects to our problem, but likely influence the polarization estimates. A more detailed treatment of the velocity distribution of the collision particles is necessary for accurate modeling of the effect we present.

Third, in our approach we assumed collisions to be dominant in determining the alignment and population of the molecular quantum states. This can safely assumed to be the case in dense gases, above the critical density. This can be contrasted to other pathways to polarization, like the Goldreich-Kylafis (GK) effect \citep{goldreich:81}, where collisions are generally thought to be unfavorable to the emergence of polarization in thermal line emission \citep{goldreich:81, lankhaar:20}. In the absence of a velocity drift between two collisional species, collisions are isotropic and work actively against the alignment of quantum states. Rather, the alignment of quantum states must be introduced through directional radiation \citep{lankhaar:20}. By comparing their interaction rates, we can estimate the ratio between collisional-to-radiative alignment to be on the order of $\sim (n_{\mathrm{H}_2} / n_{\mathrm{crit}})/\tau$ for optically thin lines, and $\sim (n_{\mathrm{H}_2} / n_{\mathrm{crit}})$ for optically thick lines. Estimating the density of the region wherefrom polarized radiation appears, as well as the optical depth of the probed transition, thus provides a natural means to distinguish between the mechanism by which the polarization has emerged. 

Last, ambipolar diffusion presupposes movement in the (charged) medium that together with the magnetic field gives rise to a Lorentz force. A definite direction in the ambipolar diffusion requires a directional flow of the gas. Turbulence on smaller scales than this flow injects an element of randomness to these motions, and thus also to the ambipolar drift. If the magnetic field is unaffected by the turbulent motions, it is conceivable that the ambipolar drift is not always perpendicular to the magnetic field. We note that therefore the perpendicular relation between the drift velocity and magnetic field is idealized. The magnetic field alignment is expected to be slightly tapered by turbulent motions.

\section{Observation of ambipolar diffusion}
We can formulate a simple relation for the polarization fraction due to collisional polarization by ambipolar diffusion by solving Eq.~(\ref{eq:pol_rad}) for low optical depth, 
\begin{align}
p_l \simeq e_q/e_i \approx -\frac{3}{4\sqrt{5}} \sqrt{\frac{(J+1)(2J+3)}{J(2J-1)}} \sigma_0^{(2)} \sin^2 \chi,
\label{eq:thin_pol}
\end{align}
where we use an analytical expression for the $w_{J'J}^{(2)}$-symbol for a $J\to J-1$ transition \citep{morris:85}. The polarization fraction decreases with $J$, the square-root factor tends to $1$ for high $J$. We note that at high optical depths, the predicted polarization fraction is consistently higher than the low optical depth estimate (see Figure~\ref{fig:pol_frac}). However, optically thin tracers are less affected by the GK effect \citep{goldreich:81}. 

As a specific example, we considered the nearby star-forming region B335 and estimated the viability of collisional polarization detection in the signal of HCO$^+$\footnote{\citet{glenn:97} set a $0.4\%$ upper limit on the polarization fractions of $J=1-0$ HCO$^+$ transitions toward the outflow lobes associated with different young protostellar systems. The polarization of such lines is expected to be the result of the GK effect because densities in these regions are too low for collisional polarization to be of importance.}. \citet{yen:18} recently constrained the ambipolar diffusion drift in this source to be $<0.3$ km~s$^{-1}$ at scales of $100$ au. They furthermore estimated the gas kinetic temperature around the protostar to vary as $T \sim 38 \ (r/100 \ \mathrm{au})^{-0.4}\ \mathrm{K}$ \citep[see also][]{evans:15} and the hydrogen number density as $n_{\mathrm{H}_2}\sim 9 \ \times 10^{6} (r/100 \ \mathrm{au})^{-2.1} \ \mathrm{cm}^{-3}$. As shown in \citet{yen:18}, optically thin H$^{13}$CO$^+$ transitions probe the midplane of the infalling envelope. Here we expect the magnetic field and ambipolar drift to be strongest. 

We focused on the H$^{13}$CO$^+$ $J=3\to 2$ transition. The collisional-to-radiative alignment ratio scales as $\sim \left(n_{\mathrm{H_2}} / n_{\mathrm{crit}}\right)/\tau$ for this optically thin species. Considering the critical density of HCO$^{+}=10^{5}$ cm$^{-3}$, and using the temperature profile to roughly estimate the thermal velocity as a function of the $r$, $v_{\mathrm{thermal}}=\sqrt{8kT/\pi \mu}\sim 0.66 \ (r/100 \ \mathrm{au})^{-0.2} \ \mathrm{km~s}^{-1}$, we can rule out contamination of the collisional polarization signal by GK polarization within some $100$ au of the protostar. 

Figure~\ref{fig:rel_align} reports the optically thin linear polarization fraction of the H$^{13}$CO$^+$ $J=3\to 2$ transition on the right-hand axis. We estimated the anisotropy parameter $b/a=0.4$ for a HCO$^+$$-$H$_2$ collision. With this parameter, the drift velocity that gives rise to a $1 \%$ polarization fraction through collisional polarization at $100$ au is $v_{\mathrm{drift}}\sim 0.1 \ \mathrm{km~s}^{-1}$. In the case of very modest anisotropy in the collisions, $b/a=0.8$, a $v_{\mathrm{drift}}\sim 0.1$ km s$^{-1}$ gives $0.2 \%$ polarization, which is just within the current ALMA detection limit. Closer to the protostar, MHD modeling predicts that the drift velocity increases more strongly than does the thermal velocity \citep{li:11, yen:18}, therefore larger polarization fractions are expected. These regions are of particular interest to the outflow-launching mechanism \citep{shu:94, blandford:82, bjerkeli:19}. However, the dynamics and magnetic field morphology become increasingly complex toward the near protostellar regions \citep{machida:08}. Our simple radiative transfer model breaks down for complex magnetic fields and flows, which means that this likely affects the polarization estimates. To model complex regions, three-dimensional polarized radiation transfer modeling using codes such as PORTAL \citep{lankhaar:20} might be used. 

\section{Conclusions}
We proposed a new method for detecting ambipolar diffusion. We showed that the velocity drift between the charged and neutral medium that characterizes ambipolar diffusion leads to a partial alignment of linear molecular ions. The molecular alignment subsequently results in partially polarized radiation emitted from these species: collisional polarization. The polarization fraction is indicative of the ambipolar drift velocity, while the polarization vectors are aligned perpendicular to the magnetic field direction. Through using optically thin species in dense regions, we can distinguish collisional polarization from polarization through the GK effect. Collisional polarization is expected in star-forming regions on scales $1-100$ au, where ambipolar diffusion is likely present, but hard to detect by other means.

We used a simple classical model for our estimates of the molecular alignment and assumed an idealized geometry for the radiative transfer of polarized radiation. More rigorous predictions can be made through employing quantum-dynamically obtained state-to-state (tensorial) cross sections of directional collisions. These cross sections can be used in conjunction with three-dimensional polarized radiation transfer \citep{lankhaar:20}; such modeling would constrain the effects of ambipolar diffusion on the emergence of polarization in molecular ions more quantitatively. 

 
%
%



\begin{acknowledgements} 
Support for this work was provided by the Swedish Research Council (VR). 
\end{acknowledgements}

\bibliography{lib}

\begin{thebibliography}{31}
\expandafter\ifx\csname natexlab\endcsname\relax\def\natexlab#1{#1}\fi

\bibitem[{Alexander {et~al.}(1977)Alexander, Dagdigian, \&
  DePristo}]{alexander:77}
Alexander, M.~H., Dagdigian, P.~J., \& DePristo, A.~E. 1977, J. Chem. Phys.,
  66, 59

\bibitem[{Andersen {et~al.}(2001)Andersen, Bartschat, \& Kessler}]{andersen:01}
Andersen, N., Bartschat, K., \& Kessler, J. 2001, Polarization, alignment, and
  orientation in atomic collisions (Springer)

\bibitem[{Bjerkeli {et~al.}(2019)Bjerkeli, Ramsey, Harsono, Calcutt,
  Kristensen, van~der Wiel, J{\o}rgensen, Muller, \& Persson}]{bjerkeli:19}
Bjerkeli, P., Ramsey, J.~P., Harsono, D., {et~al.} 2019, Astron. Astrophys.,
  631, A64

\bibitem[{Blandford \& Payne(1982)}]{blandford:82}
Blandford, R. \& Payne, D. 1982, Mon. Not. R. Astron. Soc., 199, 883

\bibitem[{Blum(1981)}]{blum:81}
Blum, K. 1981, Density Matrix Theory and Applications, Physics of atoms and
  molecules (New York: Plenum)

\bibitem[{Ciolek \& K{\"o}nigl(1998)}]{ciolek:98}
Ciolek, G.~E. \& K{\"o}nigl, A. 1998, Astrophys. J., 504, 257

\bibitem[{Degl'Innocenti \& Landolfi(2006)}]{landi:06}
Degl'Innocenti, M.~L. \& Landolfi, M. 2006, Polarization in spectral lines,
  Vol. 307 (Springer Science \& Business Media)

\bibitem[{Draine(2010)}]{draine:10}
Draine, B.~T. 2010, Physics of the interstellar and intergalactic medium
  (Princeton University Press)

\bibitem[{Ellett {et~al.}(1926)Ellett, Foote, \& Mohler}]{ellett:26}
Ellett, A., Foote, P.~D., \& Mohler, F. 1926, Physical Review, 27, 31

\bibitem[{Evans~II {et~al.}(2015)Evans~II, Di~Francesco, Lee, J{\o}rgensen,
  Choi, Myers, \& Mardones}]{evans:15}
Evans~II, N.~J., Di~Francesco, J., Lee, J.-E., {et~al.} 2015, Astrophys. J.,
  814, 22

\bibitem[{Fiedler \& Mouschovias(1993)}]{fiedler:93}
Fiedler, R.~A. \& Mouschovias, T.~C. 1993, Astrophys. J., 415, 680

\bibitem[{Flower(1999)}]{flower:99}
Flower, D. 1999, Mon. Not. R. Astron. Soc., 305, 651

\bibitem[{Friedrich \& Herschbach(1991)}]{friedrich:91}
Friedrich, B. \& Herschbach, D.~R. 1991, Nature, 353, 412

\bibitem[{Friedrich {et~al.}(1991)Friedrich, Pullman, \&
  Herschbach}]{friedrich:91b}
Friedrich, B., Pullman, D.~P., \& Herschbach, D.~R. 1991, J. Phys. Chem., 95,
  8118

\bibitem[{Glenn {et~al.}(1997)Glenn, Walker, \& Jewell}]{glenn:97}
Glenn, J., Walker, C.~K., \& Jewell, P. 1997, Astrophys. J., 479, 325

\bibitem[{Gold(1952)}]{gold:52}
Gold, T. 1952, Mon. Not. R. Astron. Soc., 112, 215

\bibitem[{Goldreich \& Kylafis(1981)}]{goldreich:81}
Goldreich, P. \& Kylafis, N.~D. 1981, Astrophys. J., 243, L75

\bibitem[{Landi~Degl'Innocenti(1984)}]{landi:84}
Landi~Degl'Innocenti, E. 1984, Sol. Phys., 91, 1

\bibitem[{Lankhaar \& Vlemmings(2020)}]{lankhaar:20}
Lankhaar, B. \& Vlemmings, W. 2020, Astron. Astrophys., 636, A14

\bibitem[{Lazarian(1997)}]{lazarian:97}
Lazarian, A. 1997, Astrophys. J., 483, 296

\bibitem[{Li {et~al.}(2011)Li, Krasnopolsky, \& Shang}]{li:11}
Li, Z.-Y., Krasnopolsky, R., \& Shang, H. 2011, Astrophys. J., 738, 180

\bibitem[{Lizano \& Shu(1989)}]{lizano:89}
Lizano, S. \& Shu, F.~H. 1989, Astrophys. J., 342, 834

\bibitem[{Machida {et~al.}(2008)Machida, Inutsuka, \& Matsumoto}]{machida:08}
Machida, M.~N., Inutsuka, S.-i., \& Matsumoto, T. 2008, Astrophys. J., 676,
  1088

\bibitem[{Mestel \& Spitzer(1956)}]{mestel:56}
Mestel, L. \& Spitzer, L. 1956, Mon. Not. R. Astron. Soc., 116, 503

\bibitem[{Morris {et~al.}(1985)Morris, Lucas, \& Omont}]{morris:85}
Morris, M., Lucas, R., \& Omont, A. 1985, Astron. Astrophys., 142, 107

\bibitem[{Mouschovias \& Ciolek(1999)}]{mouschovias:99}
Mouschovias, T.~C. \& Ciolek, G.~E. 1999, in The Origin of Stars and Planetary
  Systems (Springer), 305--340

\bibitem[{Pinto {et~al.}(2012)Pinto, Verdini, Galli, \& Velli}]{pinto:12}
Pinto, C., Verdini, A., Galli, D., \& Velli, M. 2012, Astron. Astrophys., 544,
  A66

\bibitem[{Shu {et~al.}(1994)Shu, Najita, Ostriker, Wilkin, Ruden, \&
  Lizano}]{shu:94}
Shu, F., Najita, J., Ostriker, E., {et~al.} 1994, Astrophys. J., 429, 781

\bibitem[{Sinha {et~al.}(1974)Sinha, Caldwell, \& Zare}]{sinha:74}
Sinha, M., Caldwell, C., \& Zare, R. 1974, J. Chem. Phys., 61, 491

\bibitem[{Yen {et~al.}(2018)Yen, Zhao, Koch, Krasnopolsky, Li, Ohashi, \&
  Takakuwa}]{yen:18}
Yen, H.-W., Zhao, B., Koch, P.~M., {et~al.} 2018, Astron. Astrophys., 615, A58

\bibitem[{Zhao {et~al.}(2018)Zhao, Caselli, Li, \& Krasnopolsky}]{zhao:18}
Zhao, B., Caselli, P., Li, Z.-Y., \& Krasnopolsky, R. 2018, Mon. Not. R.
  Astron. Soc., 473, 4868

\end{thebibliography}
\begin{appendix}
\section{Appended figure}
\begin{figure}[h!]
  \centering
  \includegraphics[width=0.5\textwidth]{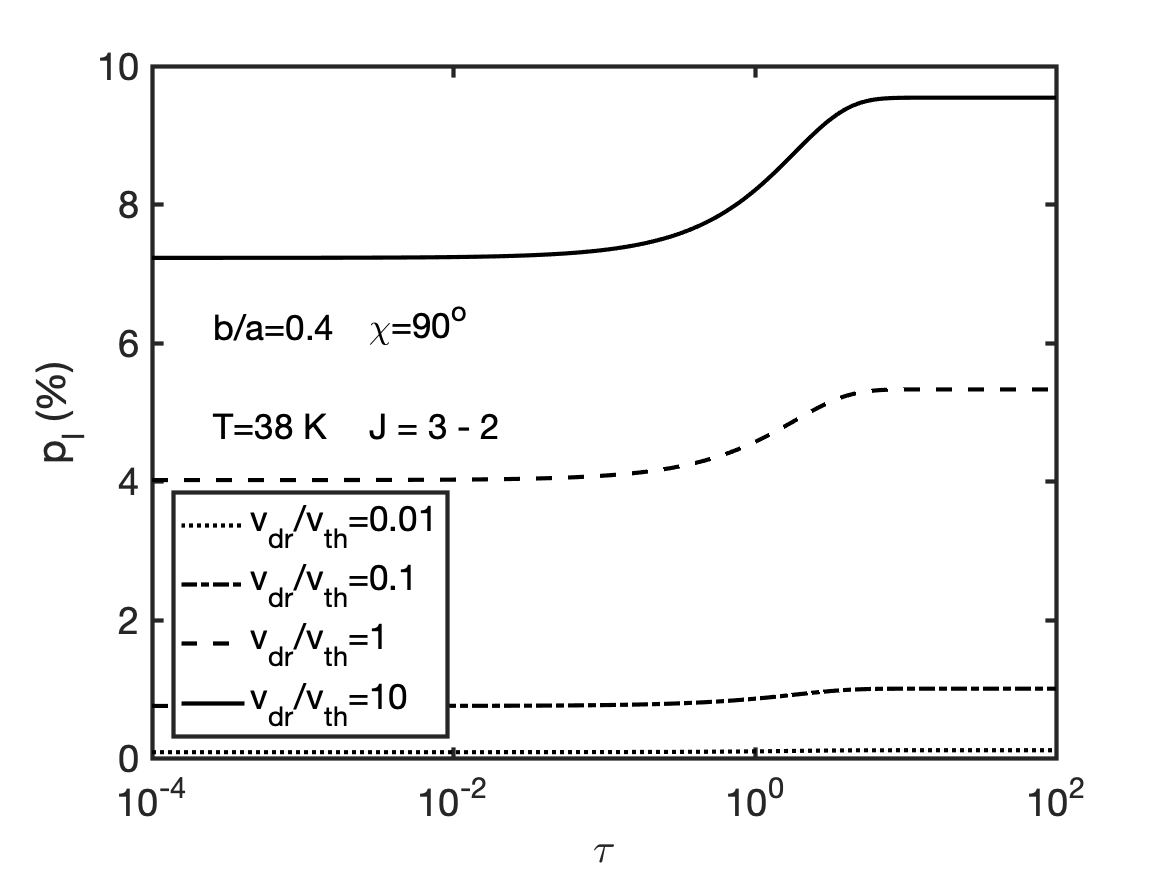}
  \caption{Polarization fraction of the $J=3-2$ transition of HCO$^+$ as a function of the optical depth. The polarization fractions resulting from different ratios of ambipolar drift velocity to thermal velocity are plotted.}
  \label{fig:pol_frac}
\end{figure}

\end{appendix}

\end{document}